\documentclass[12pt]{article}
\usepackage{graphicx}
\usepackage{epsfig}
\usepackage{caption}

\usepackage{amssymb}
\usepackage{verbatim}

\begin{document}

\title{Identification of cross and autocorrelations in time series within an approach based on Wigner eigenspectrum of random matrices}
\author{ Micha{\l} Sawa\footnote{michal\textunderscore sawa@wp.pl} \, and Dariusz Grech\footnote{dgrech@ift.uni.wroc.pl} }
\date{}

\maketitle
\begin{center}
Institute of Theoretical Physics, University of Wroc{\l}aw, Pl. M. Borna 9,\\ PL-50-204 Wroc{\l}aw, Poland\\and\\Econophysics and Time Series Analysis Group (ETSA)\\Pl. M. Borna 9, PL-50-204 Wroc{\l}aw, Poland
\end{center}

\begin{abstract}
We present an original and novel method based on random matrix approach that enables to distinguish the respective role of temporal autocorrelations inside given time series and  cross correlations between various time series. The proposed algorithm is based on properties of Wigner eigenspectrum of random matrices instead of commonly used Wishart eigenspectrum methodology. The proposed approach is then qualitatively and quantitatively applied to financial data in stocks building  WIG (Warsaw Stock Exchange Index).
\end{abstract}
$$
$$

\textbf{PACS:} 05.45.Tp, 02.60.-x, 89.20.-a, 89.75.-k, 89.65.Gh, 89.75.Fb
$$
$$
One often considers cross correlations between 1-dim time series $X^{\alpha}_i$ and $X^{\beta}_j$, where $i,j=1,...,T$ is the length of discussed data and $\alpha,\beta = 1,...,N$ mark different series. To do so the two-point simultaneous cross correlation function $C^{\alpha\beta}$ ($ -1\leq C^{\alpha\beta}\leq 1$)
  \begin{equation}
C^{\alpha \beta} = \frac{1}{T}\sum_{i,j=1}^T X^{\alpha}_i X^{\beta}_j \delta_{ij}
\end{equation}
for centered and normalized data in $X^{\alpha}$, $X^{\beta}$ is usually evaluated as a standard attempt, although cross correlations with some time lag can also be considered in a similar manner.

The very elegant way to look at global cross correlation properties between all considered series is based on random matrix (RM) approach. In this description, one calculates the spectrum of $N$ eigenvalues $\lambda_n$ ($n=1,...,N$) of $C_{N\times N}^{\alpha \beta}$ matrix (eigenspectrum), which in turn is the subject of comparison with the corresponding eigenspectrum of independent and identically distributed data $Y^{\alpha}_i$ with finite variance. The eigenspectrum $\rho(\lambda)$ of correlation matrix $M^{\alpha \beta}$, known as Wishart-Mar\v{c}enko-Pastur (WMP) spectrum \cite{9}-\cite{10}, reads for $T,N\rightarrow\infty$
\begin{equation}
\rho_{WMP}(\lambda)= \frac{Q}{2\pi \sigma}\frac{\sqrt{(\lambda_{+} - \lambda)(\lambda - \lambda_{-})}}{\lambda}
\end{equation}
where $Q=\frac{T}{N}= const$ is kept, $\sigma$ is the standard deviation of $Y^{\alpha}_i$ data and
\begin{equation}
\lambda_{\pm}=(1\pm 1/\sqrt{Q})^2
\end{equation}
are the edge values of WMP spectrum.

Since the Wishart spectrum is limited to $\lambda_{-}\leq\lambda\leq\lambda_{+}$, any deviation from this limit has the significance of cross correlation present in data. Moreover, the spread of eigenvalue spectrum of $M^{\alpha \beta}$  with respect to $\lambda_{-}\leq\lambda\leq\lambda_{+}$ tells us on the strength of such cross correlations in a system producing data $X^{\alpha}_i$ ($\alpha=1,...,N; i=1,...,T$).

This RM approach \cite{1} - \cite{4} was successfully used in econophysics and in in finance  to look for cross correlations between various one-dimensional subseries of multidimensional time series built by various stocks data (see, e.g., \cite{5} - \cite{8}).
To illustrate this idea we provide an example based on stock data taken from the main Polish stock exchange index WIG 30 in the period April 1, 2010 - Dec.\,30, 2013 what corresponds to total $T=936$ inputs of $N=26$ companies\footnote{ 26 companies of WIG 30 had long enough data history in the discussed period (stocks not included: PZU, TPE, JSW, ALR)}.

Let us define the returns
\begin{equation}
r_i=\frac{p_i-p_{i-1}}{p_{i-1}}
\end{equation}
where $p_i$ is a price of a given stock at $i$-th day which are thereafter organized into $W_{N\times T}$ matrix of centered and normalized returns.  The top part of Fig.\,1 shows the eigenspectrum of $C^{\alpha\beta}$ correlation matrix compared with the corresponding Wishart spectrum, while the bottom part of Fig.\,1 reveals the results of similar analysis done  for absolute returns $|r_i|$, i.e., the simplest transform of returns discussed in economics and econophysics.
\captionsetup[figure]{font=small,skip=0pt}

\captionsetup[figure]{font=small,skip=0pt}

\noindent
\begin{minipage}{\linewidth}
\makebox[\linewidth]{
\includegraphics[keepaspectratio=true,width=16cm,angle=0]{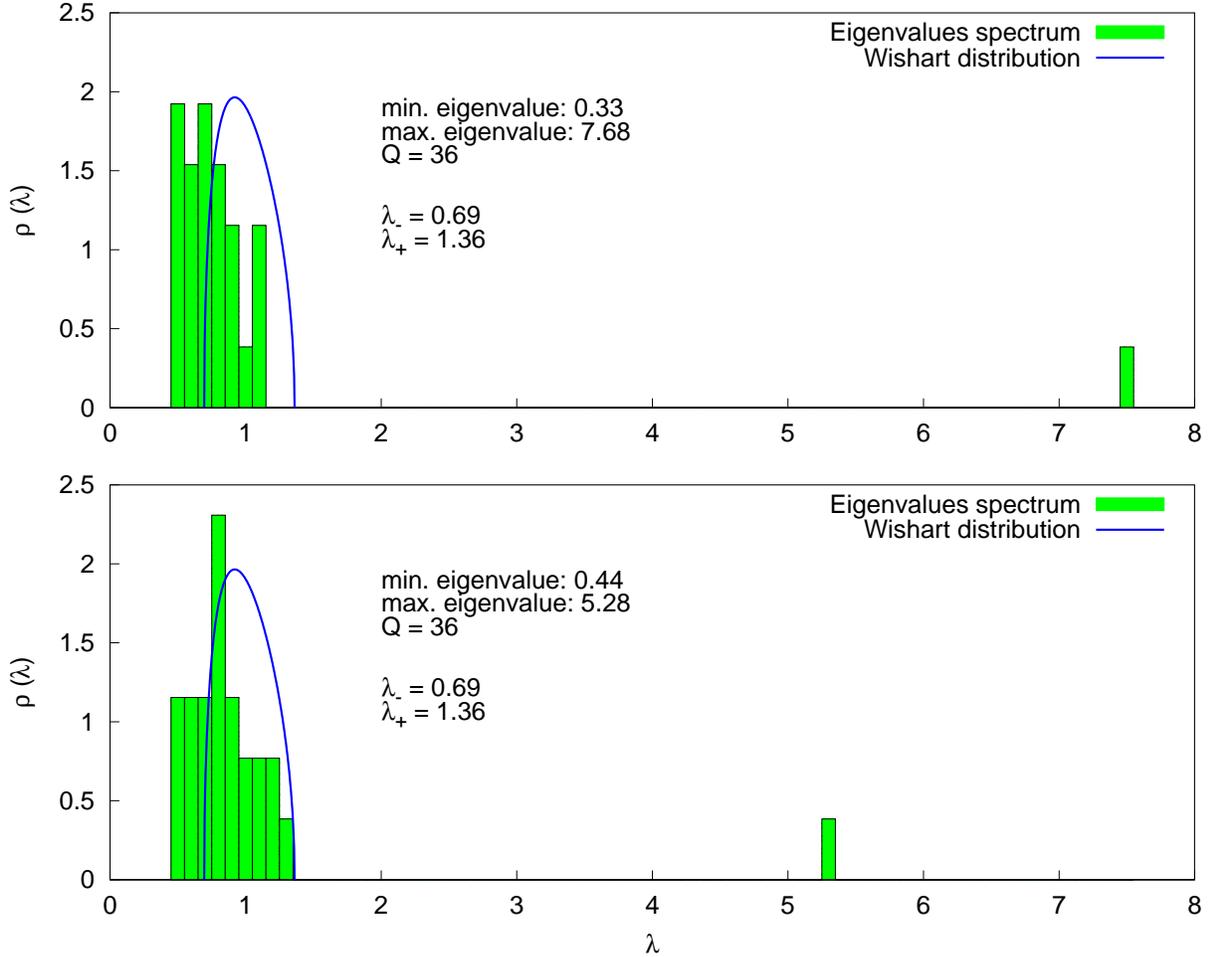}}
\captionof{figure}{Eigenspectra of correlation matrix $C(r)$ for returns (top panel) and absolute returns $C(|r|)$ (bottom) of WIG 30 data in the period April 1, 2010 - Dec.\,30, 2013. The comparison with corresponding WMP spectrum of the same $Q=T/N$ is shown (blue plot) with numerical values from Eq.(3).}
\end{minipage}

\bigskip
The deviation from Wishart spectrum is obvious in both cases thus indicating strong cross correlations between various stocks in that period.
However, it is not difficult to notice that an approach based only on investigation of differences from WMP eigenspectrum may suffer in many real situations from several shortcomings.
First, small statistics of data (particularly a small number of investigated one dimensional subseries) leads to imprecise edges of spectrum. Second, the WMP eigenspectrum method is completely robust to temporal autocorrelations in data. Thus the question arises -- to what extend the particular spectrum like the one in Fig.\,1 is a result of small number of considered time series $N$ (which theoretically should be infinite) and to what extend it is truly caused  by cross correlations present in a system.

Nevertheless, one would like to have a RM based method allowing to extract also temporal autocorrelations in data and to compare its strength with cross correlations within the same method of analysis. An important question about the mutual (relative) quantitative relation between autocorrelations in $X^{\alpha}_i$ and cross correlations between $X^{\alpha}_i$ and $X^{\beta}_j$ ($\alpha \neq \beta$) cannot be thus answered within WMP analysis. Moreover, it would be nice to have a tool that investigates not only two-point correlations but also higher order contributions to correlation features coming from averages of $n>2$ mixed random variables .

One has to be also aware that direct calculations of cross and autocorrelation properties from given data is usually a nontrivial task. It is so because:
\begin{enumerate}
\item higher orders correlation (not only two-point correlation) might be important
\item direct calculation of cross correlation (autocorrelation) function suffers from severe problems like noise present in data, possible non-stationarity and insufficient statistics (already mentioned also for WMP approach).
\end{enumerate}

We therefore propose to analyze eigenvalue spectrum of square symmetric matrices constructed of data taken from the primary $W_{N\times T}$ matrix built of $N$ time series containing $T$ data each. For the purpose of this paper we will use WIG 30 data to do so. Then we will compare the obtained eigenspectrum with the Wigner semicircle distribution \cite{11}. Note that the same approach can be easily extended to other data sets of similar form, even taken outside finance.

Let us recall that the eigenspectrum of the square, symmetric, real $N\times N$ matrix with independent $N^2$ centered entries and unit variance, known as Wigner spectrum, reads in the limit $N\rightarrow\infty$
\begin{equation}
\rho_W(\lambda)=\frac{1}{2\pi}\sqrt{4-\lambda^2}
\end{equation}

The whole analysis is similar to the one proposed by one of us (D.G.) in \cite{12}. In order to built square matrices from the original $W_{N\times T}$ matrix data we reshape it by splitting it into $[\frac{T}{m^2N}]$, ($m=1,2,...$) matrices $W_{N\times m^2N}$ with nonoverlapping entries. Then we augment these matrices one under another to form just one $W_{mN\times mN}$ matrix.

In the case of WIG 30 the original $W_{26\times936}$ matrix was reshaped into square matrix by splitting it into $6$ horizontal/time sectors of size $26\times156$, then augmenting these one under another. The final matrix obtained this way, after being symmetrized and normalized, is denoted further on as $S$ ($S_{156\times156}$ in this case). The eigenspectra of  $S_{156\times156}(r)$ and $S_{156\times156}(|r|)$ built respectively for returns $r$ and absolute returns $|r|$ are shown in Fig.2. The corresponding  Wigner eigenspectrum for uncorrelated data has also been indicated in all figures as reference.

\noindent
\begin{minipage}{\linewidth}
\makebox[\linewidth]{
\includegraphics[keepaspectratio=true,width=16cm,angle=0]{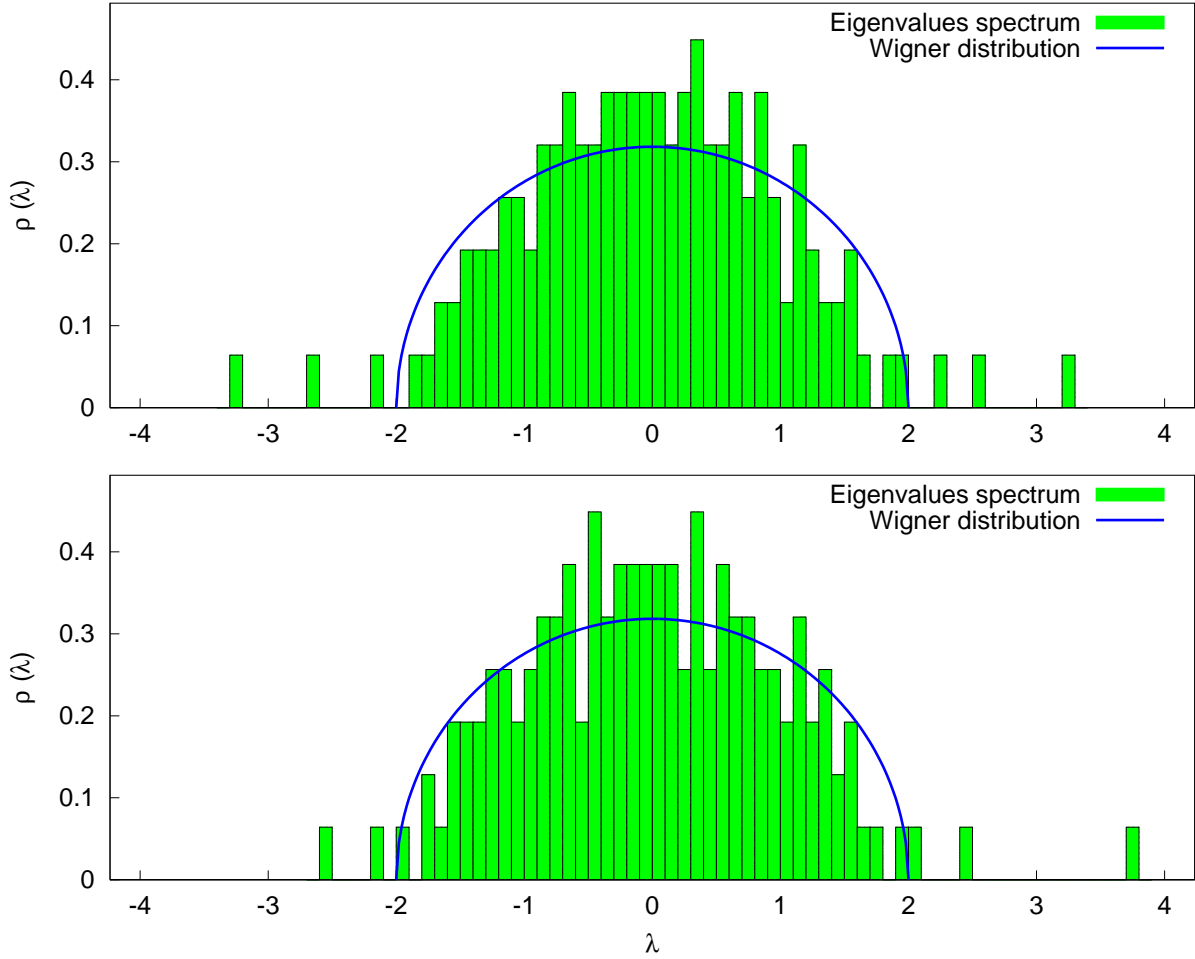}}
\captionof{figure}{Eigenspectra of $S_{156\times 156}(r)$ (top) and $S_{156\times 156}(|r|)$ (bottom) matrices constructed from $W_{26\times 936}(r)$ and $W_{26\times 936}(|r|)$ respective data of WIG 30. The comparison with corresponding Wigner eigenspectrum of uncorrelated data is made (blue curve).}
\end{minipage}

\bigskip
The original $W_{26\times 936}$ matrix can be 'reshaped' also into a set of smaller square matrices by repeating the procedure described above for its smaller parts - horizontal/time sectors. For example considering sectors $26\times104$ one may produce $9$ matrices $52\times52$. The averaged eigenspectrum of these for $r$ and $|r|$ are presented in Fig.\,3.

\noindent
\begin{minipage}{\linewidth}
\makebox[\linewidth]{
\includegraphics[keepaspectratio=true,width=16cm,angle=0]{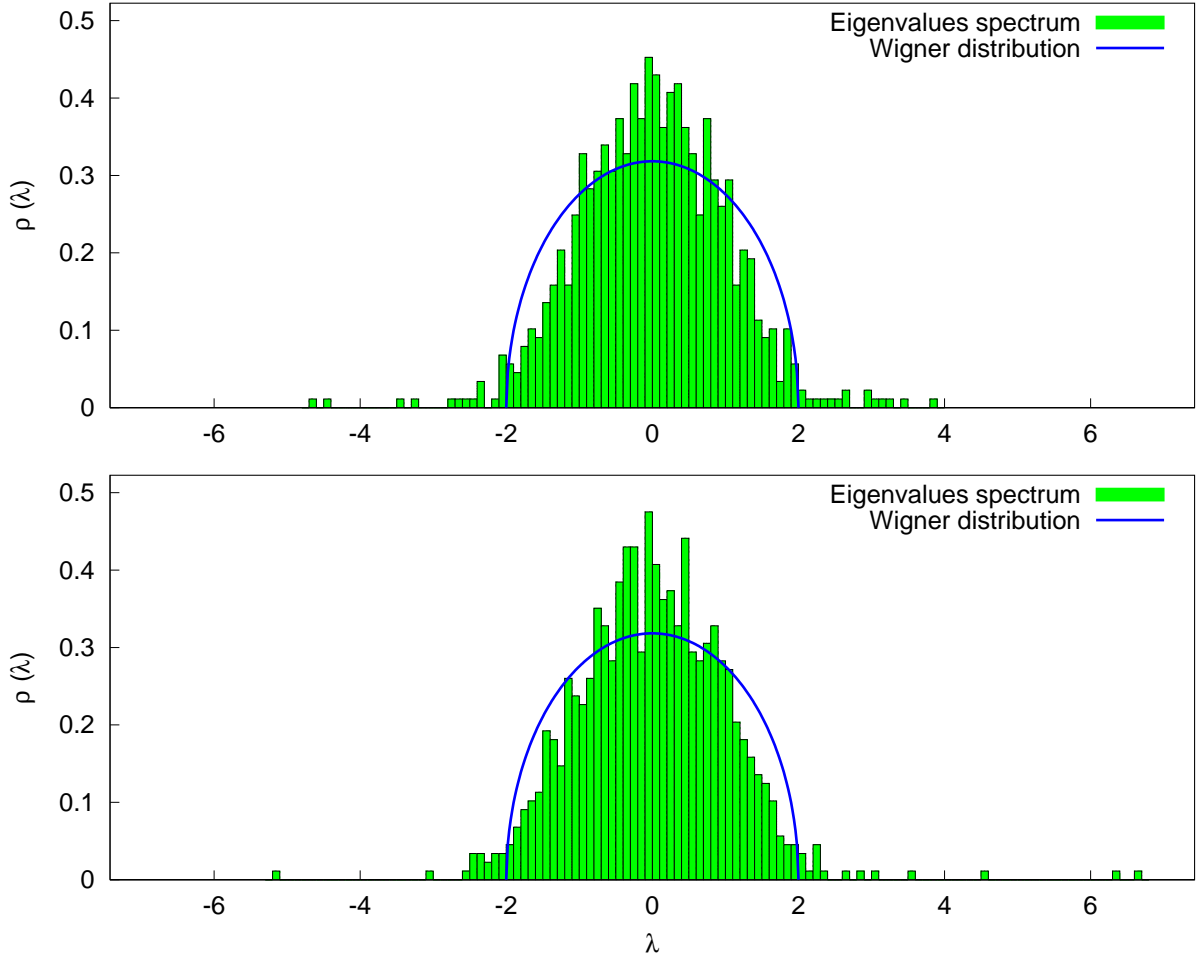}}
\captionof{figure}{Averaged eigenspectra of matrices $S_{52\times 52}(r)$ (top) and $S_{52\times 52}(|r|)$ (bottom) constructed from $W_{26\times 936}(r)$ and $W_{26\times 936}(|r|)$ data of WIG 30 respectively. The Wigner semicircle for uncorrelated data of the same size is drawn for comparison (blue curve).}
\end{minipage}

\bigskip
By folding up the horizontal sectors in various ways, one obtains a more abundant statistics from the available data (compare Fig.\,2 and Fig.\,3). It is clearly visible that the spectrum of the new-built square matrices in the examples presented above exceeds the theoretical range expected for Wigner eigenspectrum. It differs also from the averaged eigenspectra of matrices $S_0$ obtained by reshaping completely shuffled  $W_{N\times T}$ data to remove all cross and autocorrelations. This point is well clarified in Fig.\,4.

\noindent
\begin{minipage}{\linewidth}
\makebox[\linewidth]{
\includegraphics[keepaspectratio=true,width=16cm,angle=0]{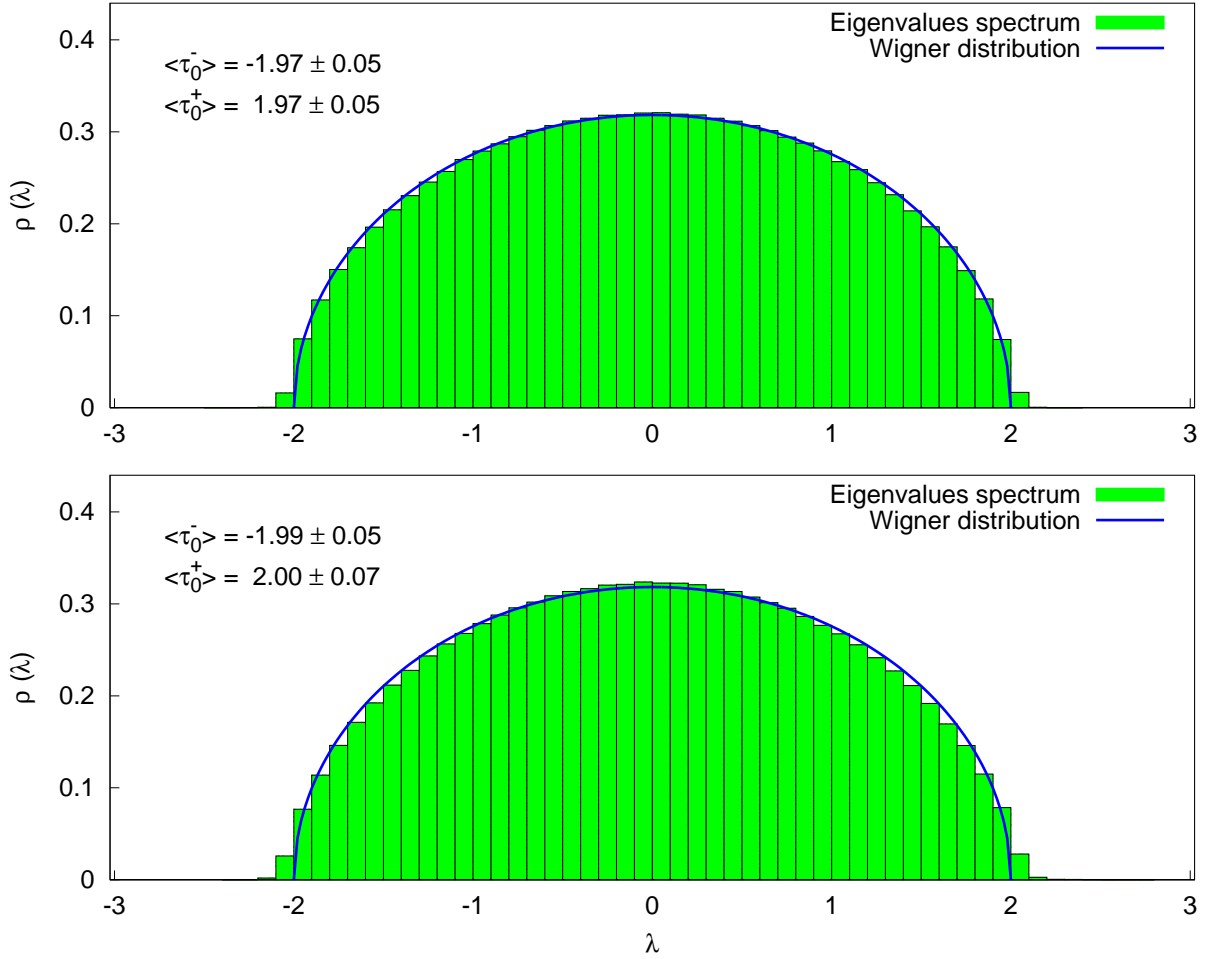}}
\captionof{figure}{Averaged eigenspectra of $S_0(r)$ (top) and $S_0(|r|)$ (bottom) matrices constructed from $10^4$ shuffles of $W_{26\times 936}(r)$ and $W_{26\times 936}(|r|)$ data matrix of WIG 30 respectively. The very good correspondence with Wigner distribution (blue plot) is indicated.}
\end{minipage}

\bigskip
The eigenspectrum of $S_0 (r)$ matrix  has the averaged tail edges positioned at: $\langle\tau_0^+\rangle(r) = 1.97\pm 0.05$ and $\langle\tau_0^-\rangle(r) = -1.97\pm 0.05$, where superscript $\pm$ corresponds to right and left tail respectively. The corresponding results for eigenspectrum of $S_0(|r|)$ matrix built for absolute returns read: $\langle\tau_0^+\rangle(|r|) = 2.00\pm 0.07$ and $\langle\tau_0^-\rangle(|r|) = -1.99\pm 0.05$. In all cases the mean and the standard deviation is taken from the ensemble of $10^4$ matrices eigenspectra since the process of shuffling, reshaping and calculating eigenvalues was repeated here $10^4$ times. The spectra in Figs.\,2,\,3 compared with Fig.\,4 clearly indicate an existence of correlation (autocorrelation and/or cross correlation) in analyzed signals.

Now, we may ask whether within RM method one is able to distinguish the role and respective weights of cross and autocorrelations present in signals. Note that the answer to such question was negative in case of standard $C^{\alpha\beta}$ correlation approach based on WMP eigenspectrum.
It turns out that it is possible to exhaust such information by applying diversified shuffling methods to all signals.  In order to kill cross correlations between different companies (rows in $W_{N\times T}$ matrix), preserving however autocorrelations (of all orders), we make random 'cyclic shifting' of data in rows of original matrix $W_{N\times T}$ before reshaping it into $S_{156\times156}$ one with no cross correlations present - denoted here further on by $S_{ac}$). The detailed procedure like this is as follows.

A natural number $n_{\alpha}$,  ($1\leq n_{\alpha}\leq T$,  $\alpha = 1,...,N$) is chosen at random from discrete uniform distribution separately for each row of $W_{N\times T}$ matrix and a new  matrix $W^{(cs)}_{N\times T}$ is formed with cyclically shifted rows -- the $\alpha$-th row $(w_{\alpha 1}, w_{\alpha 2},..., w_{\alpha T})$ is replaced by $(w_{\alpha n_{\alpha}}, w_{\alpha n_{\alpha}+1},...,w_{\alpha T}, w_{\alpha 1}, w_{\alpha 2},..., w_{\alpha n_{\alpha}-1})$. Then $W^{(cs)}_{N\times T}$ matrix is reshaped into a square matrix and symmetrized as before. We calculate its spectrum and the process is repeated $10^4$ times. It results in spectra (averaged eigenspectrum) presented in Fig.\,5.

On the other hand, to kill autocorrelations preserving however cross correlations (of any order), we perform random 'shuffling' of columns of the original data matrix reshaping resultant matrices into another $S_{156\times156}$ ones denoted further on by $S_{cc}$. Due to the way the shuffling was done in this case cross corelations are still preserved. The corresponding averaged eigenspectrum in this case after $10^4$ repetitions is presented in Fig.\,6.  The average position of the tail edge of eigenvalue spectra is marked in a similar manner as before, i.e., $\langle\tau^{\pm}_{cc}\rangle$ in case of killed autocorrelations (cross correlations are left only), and $\langle\tau^{\pm}_{ac}\rangle$ in case of killed cross correlations (autocorrelations being left).

\noindent
\begin{minipage}{\linewidth}
\makebox[\linewidth]{
\includegraphics[keepaspectratio=true,width=16cm,angle=0]{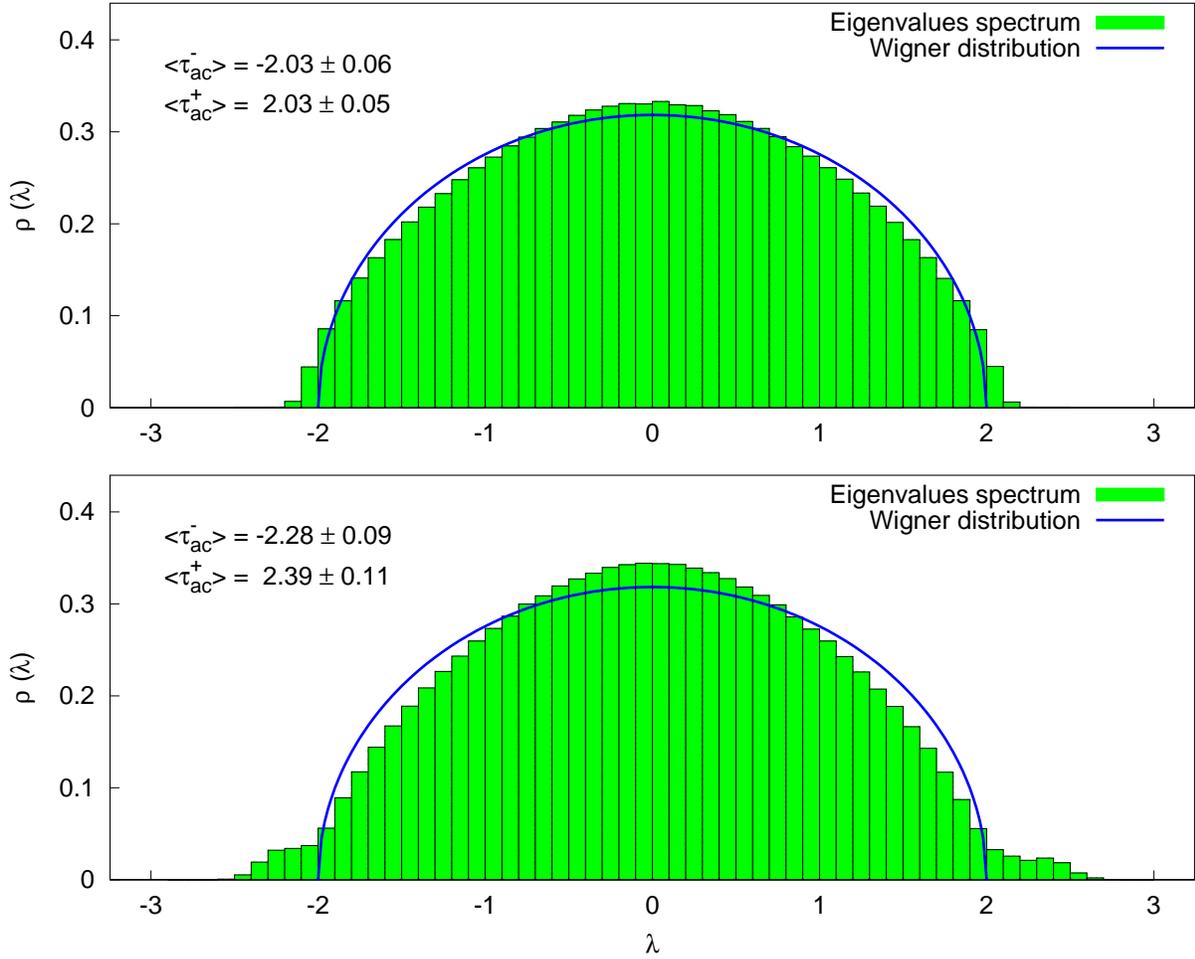}}
\captionof{figure}{Averaged eigenspectra of $S_{ac}(r)$ matrix (top) with 'killed' cross correlations (only autocorrelations are left in signal) extracted from $W_{26\times 936}(r)$ of WIG30 data. The same for absolute returns is shown for $S_{ac}(|r|)$ matrix (bottom). Wigner spectrum is shown for comparison (blue curve).}
\end{minipage}

\noindent
\begin{minipage}{\linewidth}
\makebox[\linewidth]{
\includegraphics[keepaspectratio=true,width=16cm,angle=0]{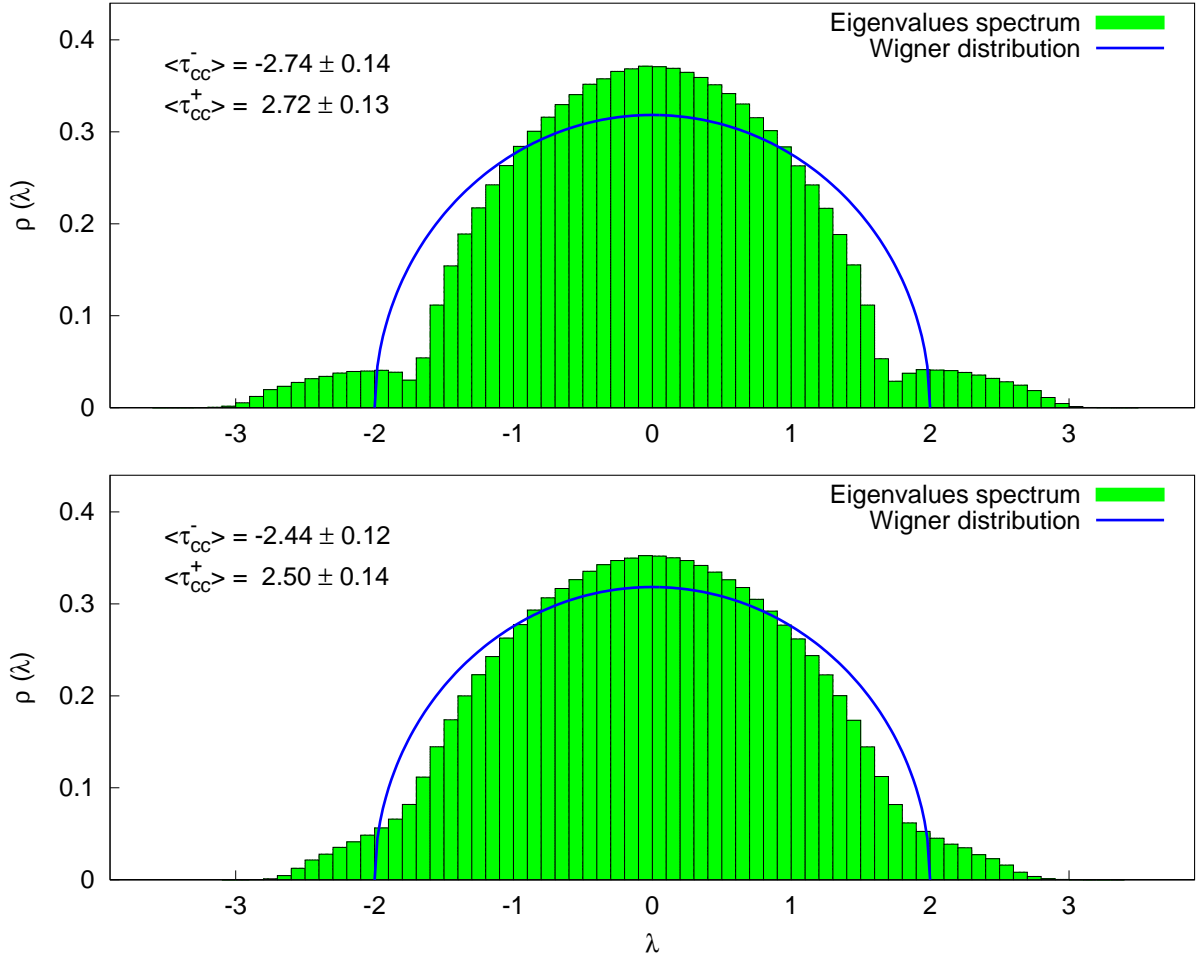}}
\captionof{figure}{Same as in Fig.5 but for 'killed' autocorrelations in signal (only cross correlations are present).}
\end{minipage}

\bigskip
The noticed differences in average tail lengths shown in Fig.\,5 (with autocorrelation extracted) and Fig.\,4 (for the shuffled signal with no correlation at all) indicate detection of very weak autocorrelations in returns and somehow more signicant autocorrelations in absolute returns (both of any order). On the other hand, the comparison of results presented in Fig.\,6 (cross correlation extracted) with those in Fig.\,4 indicate detection of cross correlations between time series. Thus the presented analysis is not only able to detect correlations in multidimensional data but offers also possibility to detect separately autocorrelations (Fig.\,5) and cross correlations(Fig.\,6). Moreover, the differences in mean tail lengths shown in consecutive figures if compared with reference plot of Wigner semicircle for uncorrelated matrix entries may be exploited to estimate quantitatively relative weights of cross correlations and autocorrelations.

 The predominance of cross correlations over autocorrelations in returns $(r)$ is visible as $|\langle\tau^{\pm}_{cc}\rangle(r)| \gg |\langle\tau^{\pm}_{ac}\rangle(r)|$ if one compares the average lengths of eigenspectrum distributions for $S_{ac}(r)$ and $S_{cc}(r)$. This difference does not manifest as much for absolute returns $|r|$ since here $|\langle\tau^{\pm}_{cc}\rangle(|r|)| \gtrsim |\langle\tau^{\pm}_{ac}\rangle(|r|)|$ (compare bottom panels of Fig.\,5 and 6).
The relative strength of cross correlations and autocorrelations can be estimated introducing the subsequent ratios defined in a straightforward manner

\begin{equation}
\Delta_{{cc}/{ac}}(r) = \frac{[\langle\tau^{+}_{cc}\rangle(r) - \langle\tau^{-}_{cc}\rangle(r)]-[\langle\tau^{+}_{0}\rangle(r)- \langle\tau^{-}_{0}\rangle(r)]}{[\langle\tau^{+}_{ac}\rangle(r) - \langle\tau^{-}_{ac}\rangle(r)]-[\langle\tau^{+}_{0}\rangle(r)- \langle\tau^{-}_{0}\rangle(r)]}
\end{equation}
in case of returns, and

\begin{equation}
\Delta_{{cc}/{ac}}(|r|) = \frac{[\langle\tau^{+}_{cc}\rangle(|r|) - \langle\tau^{-}_{cc}\rangle(|r|)]-[\langle\tau^{+}_{0}\rangle(|r|)- \langle\tau^{-}_{0}\rangle(|r|)]}{[\langle\tau^{+}_{ac}\rangle(|r|) - \langle\tau^{-}_{ac}\rangle(|r|)]-[\langle\tau^{+}_{0}\rangle(|r|)- \langle\tau^{-}_{0}\rangle(|r|)]}
\end{equation}
for  absolute returns $|r|$.

Very similar analysis allows to compare quantitatively the autocorrelation levels  between returns $(r)$ and absolute returns $|r|$
\begin{equation}
\delta_{ac}(|r|/r) = \frac{[\langle\tau^+_{ac}\rangle(|r|) - \langle\tau^-_{ac}\rangle(|r|)]-[\langle\tau^+_{0}\rangle(|r|)- \langle\tau^-_{0}\rangle(|r|)]}{[\langle\tau^+_{ac}\rangle(r) - \langle\tau^-_{ac}\rangle(r)]-[\langle\tau^+_{0}\rangle(r)- \langle\tau^-_{0}\rangle(r)]}
\end{equation}
and analogously, between change of cross correlation levels while we transfer from returns to absolute returns
\begin{equation}
\delta_{cc}(|r|/r) = \frac{[\langle\tau^+_{cc}\rangle(|r|) - \langle\tau^-_{cc}\rangle(|r|)]-[\langle\tau^+_{0}\rangle(|r|)- \langle\tau^-_{0}\rangle(|r|)]}{[\langle\tau^+_{cc}\rangle(r) - \langle\tau^-_{cc}\rangle(r)]-[\langle\tau^+_{0}\rangle(r)- \langle\tau^-_{0}\rangle(r)]}
\end{equation}
Substituting numerical values of tail positions for eigenspectra of the above generated matrices we find $\Delta_{{cc}/{ac}}(r)\approx 12.67$ and $\Delta_{{cc}/{ac}}(|r|)\approx 1.40$. Simultaneously, the relative strength of autocorrelations between $|r|$ and $r$ increases to the level $\delta_{ac}(|r|/r)\approx 5.67$ while the strength of cross correlations drops down for $|r|$ comparing it with $r$ since for the latter ratio one finds $\delta_{cc}(|r|/r)\approx 0.63$.

Concluding, we state that
 RM approach based on Wigner spectrum analysis can be used to compare qualitatively and quantitatively different forms of correlations in multidimensional data. In our example of Warsaw Stock Exchange data, the role of autocorrelations increases and the role of cross correlations decreases when one proceeds from returns $(r)$ to absolute returns $|r|$ (thus revealing the importance of sign in returns). The approach considering the average tail lengths of probability distribution we presented in this article (instead
of cumulative distributions of matrix eigenspectra) seems to be statistically more
reliable since it eliminates large fluctuations from eigenspectrum. This is why we adopted the first one.

Since in this RM based approach higher orders of correlations as well as the influence of short term memory is included, we should not be surprised that very small autocorrelation effect in primary data was observed (see Fig.\,5 in comparison with Fig.\,4). This effect increases by about six times when one moves from $r$ to $|r|$.
Simultaneously the cross correlation level in between series is sensitive to the presence of sign in returns in an opposite direction. The cross correlations strength between absolute returns  is about $40\%$ smaller than for  returns confirming an importance of influence of sign in price change on cross correlation magnitude.

Note that in the presented analysis all quantitative results take automatically into account also higher correlation orders. Therefore they are more general than the outcomes of the standard approach when only two-point correlation function is considered.
Finally, it is worth emphasizing  that this kind of analysis can be easy extended to investigate mutual relationship between arbitrary data in multidimensional time series of any origin.

\end{document}